\documentclass[%
 reprint,
superscriptaddress,
 amsmath,amssymb,
pra,
]{revtex4-1}

\usepackage{graphicx}
\usepackage{dcolumn}
\usepackage{bm}
\usepackage{braket}

\begin{document}

\preprint{APS/123-QED}

\title{A Turn-Key Diode-Pumped All-Fiber Broadband Polarization-Entangled Photon Source}

\author{Changjia Chen}
\email{changjia.chen@mail.utoronto.ca}
\author{Arash Riazi}
\author{Eric Y. Zhu}
\affiliation{Dept.of Electrical and Computer Engineering, University of Toronto, Toronto, M5S 3G4, Canada}
\author{Mili Ng}
\affiliation{Ozoptics Ltd., 219 Westbrook Rd, Ottawa, K0A 1L0, Canada}
\author{Alexey V. Gladyshev}
\affiliation{Fiber Optics Research Center, Russian Academy of Sciences, 119333 Moscow, Russia}
\author{Peter G. Kazansky}
\affiliation{Optoelectronics Research Centre, University of Southampton, Southampton, United Kingdom}
\author{Li Qian}
\affiliation{Dept.of Electrical and Computer Engineering, University of Toronto, Toronto, M5S 3G4, Canada}
\affiliation{National Engineering Laboratory for Fiber Optic Sensor Technology, Wuhan University of Technology, Wuhan 430070, China}

\date{\today}

\begin{abstract}
In this letter, we report a compact, low-power laser diode-pumped, all-fiber polarization-entangled photon pair source based on periodically-poled silica fiber technology. The all-fiber source offers room-temperature, alignment-free, turn-key operation, with low power consumption, and is packaged into a fanless, portable enclosure. It features a broad biphoton spectrum of more than 100nm with a concurrence that is greater than 0.96 for polarization entanglement. The source is stable over at least 10 hours of continuous operation, achieving coincidence-to-accidental ratios of more than 2000 consistently over this time period.  

\end{abstract}

\maketitle


Practical applications of quantum entanglement-based photonic technologies such as quantum interferometry\cite{giovannetti2004quantum} and long-distance quantum key distribution\cite{yin2017satellite,wengerowsky2018field} often require robust and high performance entangled photon source. Polarization entanglement is one of the most favourable degrees of freedom exploited for entanglement, due to the ease of manipulation, generation, and detection of polarization states. Therefore it has been widely applied in free-space quantum communication\cite{bedington2017progress, yin2017satellite2, ursin2007entanglement} and can also be used in fiber-based metropolitan-scale entanglement distribution\cite{wengerowsky2018field} and quantum sensing\cite{simon2017quantum}. Reliable and robust sources of polarization-entangled photon pairs are highly sought after.

Progress of practical and commercial polarization entanglement sources so far has been mostly made in the design and optimization of nonlinear crystals or waveguides\cite{steinlechner2012high, poppe2004practical, durak2016next, steinlechner2016development, liao2017satellite, tang2016generation}. These sources, while being efficient and bright in the generation of photon pairs, still require tedious optical alignment, employ free-space or waveguide coupling optics, and thus require expensive rugged packaging to withstand harsher conditions of in-field operation. In contrast, fiber-based sources employing mature telecommunication components can replace expensive bulk optics with simple fiber connectors or splicing, eliminating beam alignment, and are thus more robust against environmental changes. They are also compatible with existing low-loss telecom fiber links, facilitating long-distance entanglement distribution. Though polarization-entangled photon sources based on dispersion-shifted fibers\cite{lee2006generation}, polarization-maintaining fibers\cite{fan2007bright,fang2014polarization,zhu2016fiber}, and photonic crystal fibers\cite{fulconis2007nonclassical} have been investigated in the past decades, due to the lack of second-order nonlinearity and difficulty in phase-matching, they either do not operate in the telecom wavelength region, or require cryogenic cooling to reduce Raman noise. In addition, temporal compensation or erasure of distinguishability through interferometry is always required in these types of fiber-based sources to achieve high-quality entanglement, which not only greatly increases the complexity of the construction, but also requires the use of free-space optics in most cases.

In contrast to the biphoton sources based on the third-order nonlinearity of optical fiber, periodically-poled silica fiber (PPSF) \cite{helt2009proposal} uses the induced second-order nonlinearity in poled fiber to generate polarization-entangled photon pairs \cite{zhu2012direct} based on type-II spontaneous parametric down conversion (SPDC). The negligible group birefringence of the PPSF ensures high-quality and broadband polarization entanglement without compensation\cite{Chen2017}, making it particularly suitable for applications such as broadband quantum sensing\cite{kaiser2018quantum,riazi2018quantum}, multi-channel quantum communication and information processing\cite{aktas2016entanglement,li2016multiplexed,zhu2015multi, wengerowsky2018entanglement, kues2017chip}. 

Based on the PPSF technology, here we present a turn-key, all-fiber, telecom-band polarization-entanglement source by incorporating a low-cost fiber-pigtailed pump diode directly spliced to our PPSF. The compensation-free nature \cite{Chen2017} of the PPSF source obviates the need for temporal compensators or interferometers. Combined with off-the-shelf fiber-pigtailed filter components, our alignment-free source is packaged into a plug-and-play enclosure. It has to our knowledge the simplest design of all polarization-entangled sources reported so far in the literature. 

A detailed description of what is inside the enclosure is as follows: A fiber-coupled Fabry-Perot cw laser diode module (OZ Optics, OZ-1000) is used as a pump. It is mounted on a thermal-electric cooler for temperature control. Its output wavelength is stabilized to 782.90$\pm$0.05nm to match the type-II peak of the PPSF. The pump diode has a linewidth of less than 0.05 nm, and a fiber-coupled output power adjustable from 1 mW to 30 mW. The pump outputs in one linear polarization mode, which is aligned to the fast axis of the single-mode polarization-maintaining fiber (PMF) pigtail. 
\begin{figure}
\centering
\includegraphics[width=8.3cm]{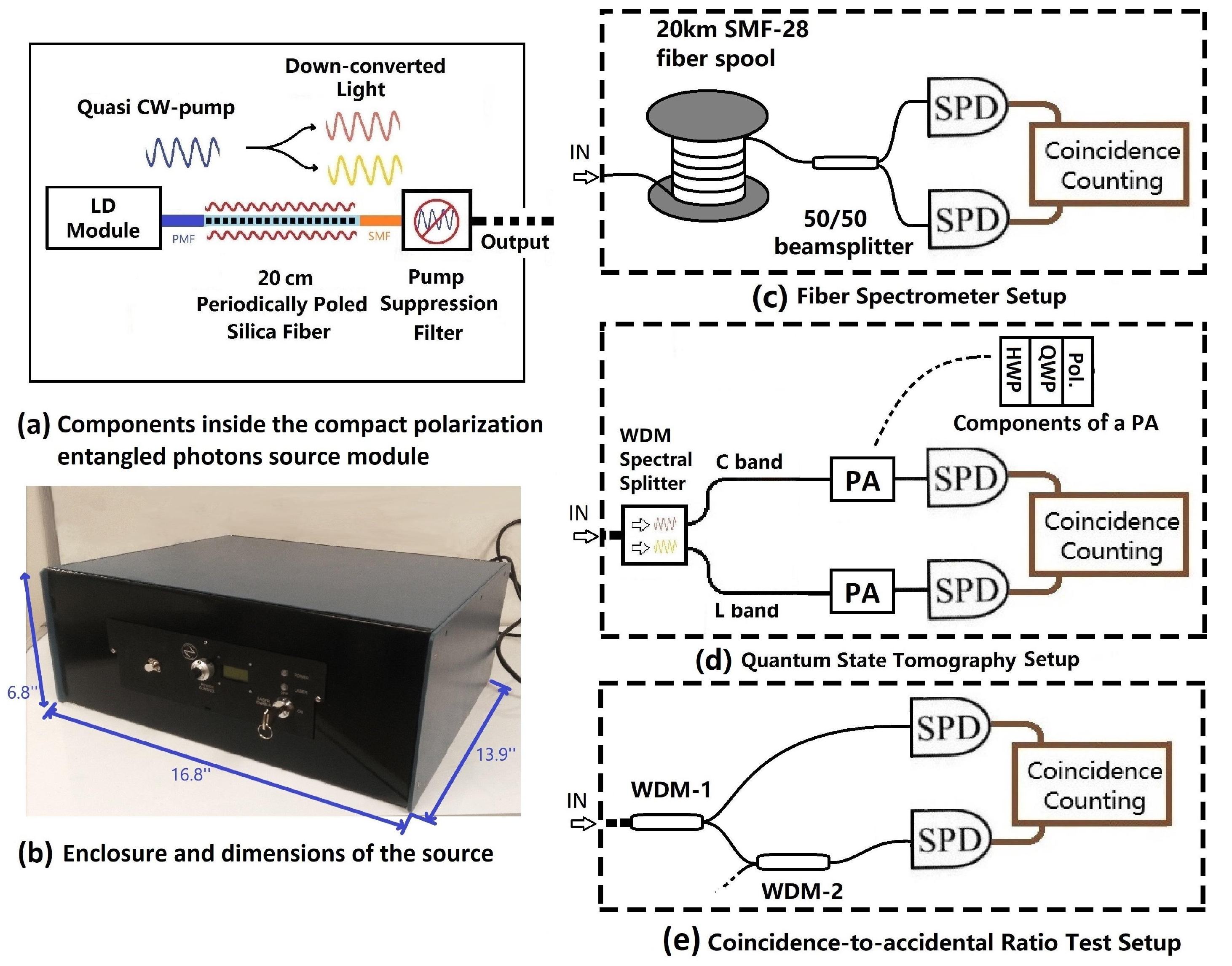}
\caption{(a) The PPSF-based polarization entangled photon pair source module includes: a polarization-maintaining fiber pigtailed laser diode, a 20cm long PPSF, and a set of pump suppression filters. (b) The setup in (a) and a power supply can be easily fit into a 3U enclosure. (c) Fiber spectrometer. (d) Quantum state tomography(QST) setup for characterization of broadband polarization entanglement. The QST setup includes wavelength splitter (L/C band WDM), polarization analyzers(PAs) and a pair of single photon detectors (SPDs). Each PA consists of an achromatic half-(HWP) and quarter wave plate(QWP) and a polarizer (Pol).  (e) Coincidence-to-accidental stability test setup. WDM-1/2 is a pair of bandpass filters centering at 1554.95/1577.10nm with 1.1nm FWHM bandwidth for seperation of entangled photons.}
\label{fig:setup}
\end{figure}
The 20-cm long PPSF is packaged with a low-cost resistive strip heater and is temperature controlled to 34$\pm$0.3$^o$C. This temperature can be set higher or lower depending on ambient temperature range. A temperature deviation of $\pm$0.3$^o$C results in a phase-matching wavelength variation of $\pm$ 0.03nm. The PPSF is spliced to the PM fiber pigtail of the pump such that both their slow and fast axes are aligned. Launching pump light into the PPSF along the fast axis is optimal for the generation of polarization entangled photon pairs via the type-II SPDC process \cite{zhu2012direct}). Note, due to the spectral separation of type-0 and type-II peaks (see Ref.\cite{Chen:18, zhu2010measurement}), precise polarization alignment is not required for the PMF-PPSF splice, nor for the PMF pigtail of the laser diode, as misalignment only leads to a drop in pair generation efficiency but not in entanglement quality. At the output end of the PPSF, it is spliced to a single-mode-fiber (SMF) with negligible birefringence. It is then followed by conventional off-the-shelf fiber-pigtailed filters for pump and noise suppression. The PPSF's degeneracy wavelength is set at 1565.80$\pm$0.10nm, approximately at the boundary between the C and L band of the telecom wavelength. The biphotons can then be conveniently split in wavelength with an off-the-shelf C/L band wavelength-division multiplexer, also referred to as a C/L band splitter. 

As our entangled source is composed of very few components, most of which are passive, the resulting package is extremely stable, compact, and robust.  Because of the all-fiber structure, no beam or polarization alignment is required. All components, including a low-power-consumption power supply, can easily be fit into a standard 13.9"L$\times$16.8"W$\times$6.8"H (3U) enclosure, as illustrated in Fig.\ref{fig:setup}(b). As there is still ample empty space inside the enclosure, a smaller enclosure, for example, with a height of 1U, can also be used.

The bipartite polarization entanglement state generated by type-II SPDC can be written as\cite{humble2007spectral}:
\begin{align}
\ket{\Phi}\propto\iint\left[f_-(\omega_s,\omega_i)\ket{H_sV_i}+f_+(\omega_s,\omega_i)\ket{V_sH_i}\right]\mathrm{d}\omega_s\mathrm{d}\omega_i\label{Eqn:state}
\end{align}
The subscripts \textit{s,i} denote signal and idler photons respectively. The $f_-$ and $f_+$ are the frequency-dependent complex amplitudes. Due to the low birefringence property of the PPSF, the absolute amplitudes $|f_-|$ and $|f_+|$ are spectrally similar\cite{humble2007spectral}. Previously we have also demonstrated the direct polarization entanglement in PPSF with narrowband filtering and theoretically show that their frequency-dependent phase difference, which indicate the amount of walk-off between two states, is negligible over the whole bandwidth\cite{Chen2017}.  Hence no compensation or erasure of distinguishability is needed at the PPSF output. 

When the bandwidth of generated entangled biphotons is narrow, the phase and amplitude difference between $f_-(\omega_s,\omega_i)$ and $f_+(\omega_s,\omega_i)$ in Eq.(\ref{Eqn:state}) is essentially zero, due to the negligible group birefringence of our source. Even in the case of broadband generation over several THz, the state described by Eq.(\ref{Eqn:state}) for our PPSF source still maintains high polarizataion-entanglement quality due to its low group birefringence. To demonstrate the genuine broadband nature of our source, we first characterized the spectral property of the source using a fiber spectrometer\cite{zhu2015self,avenhaus2009fiber}. 
\begin{figure}
\centering
\includegraphics[width=8.5cm]{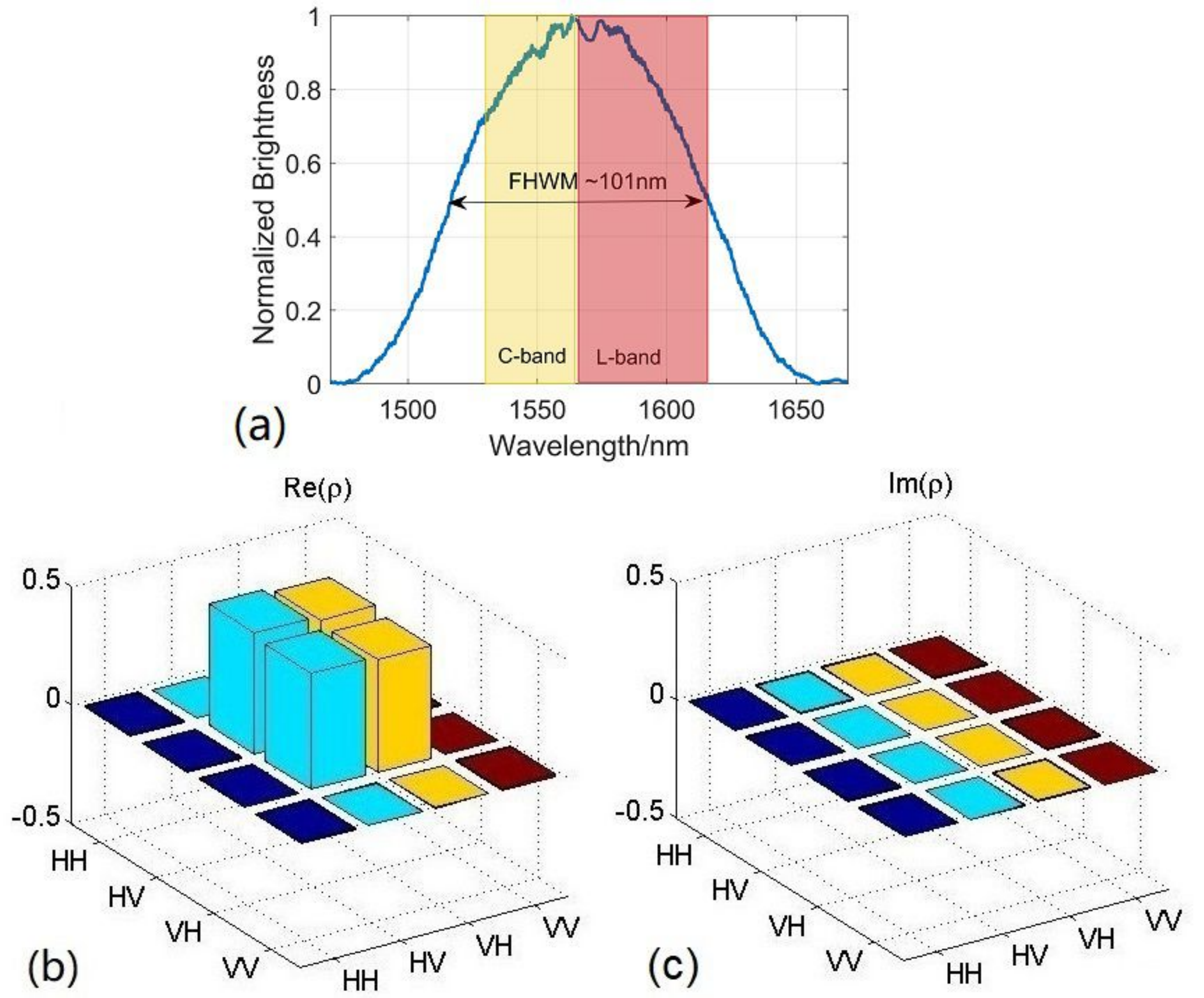}
\caption{(a) The biphoton spectrum of the PPSF-based compact entanglement source measured from a fiber spectrometer. Accidentals have been subtracted from the spectrum. The C/L band filter edge is also indicated on the graph to show that the signal and idler can be conveniently separated with a standard C/L band filter. (b) and (c): Real and imaginary part of the reduced density matrix (traced over frequency) obtained by QST on signal and idler separated by C/L band filter. The bandwidth covered by the signal and idler is ~70nm. A concurrence of 0.966$\pm$0.015 and and a fidelity to $\ket{\Psi^{+}}$ of 98.1$\pm$0.8$\%$ is obtained, without subtraction of accidentals.}
\label{fig:densitymatrix}
\end{figure}
The spectrometer setup is shown in Fig.\ref{fig:setup}(c). The biphotons generated by our compact source are temporally dispersed by a 20 km spool of Corning SMF28 fiber. The spool has a known dispersion that was obtained from a previous calibration\cite{zhu2015self}, and the relative delays of a signal photon with respect to its idler correspond in a one-to-one fashion to their spectral separation, allowing us to measure the spectrum of the biphotons. The dispersive medium is then followed by a 50/50 non-polarizing fiber beamsplitter. Note that compared to the finite bandwidth spectral filters\cite{zhu2015self}, the beamsplitter scheme can cover the whole spectrum and thus gives a more complete spectral information. The temporal-to-spectral mapping of the coincidence measurement results in a biphoton spectrum with accidentals subtracted as shown in Fig.\ref{fig:densitymatrix}(a). A FWHM bandwidth of 101nm is thus determined. 

Next we performed QST in polarization degree of freedom (Fig.\ref{fig:setup}(d)) using a commercial broadband C/L band splitter to separate the signal and idler into C (1530-1565nm) and L (1565-1615nm) bands. The effective spectral detection window is $\sim$35 nm (corresponding to the C band bandwidth, which is narrower). A pair of polarization analyzers and single photon detectors (IDQ-220, 20$\%$ detection efficiency at 1550nm) are used for coincidence measurements. As the C/L band splitter does not have its boundary exactly matching the degeneracy wavelength, some degradation in the measured concurrence/fidelity results. In spite of this, the measured density matrix (Fig.\ref{fig:densitymatrix}(b) and (c)) yields a high concurrence of 0.966$\pm$0.015 and a high fidelity to $\ket{\Psi^+}$ of 98.1$\pm$0.8$\%$ without subtraction of the background over a broad bandwidth.  
\begin{figure}[t]
\centering
\includegraphics[width=8.7cm]{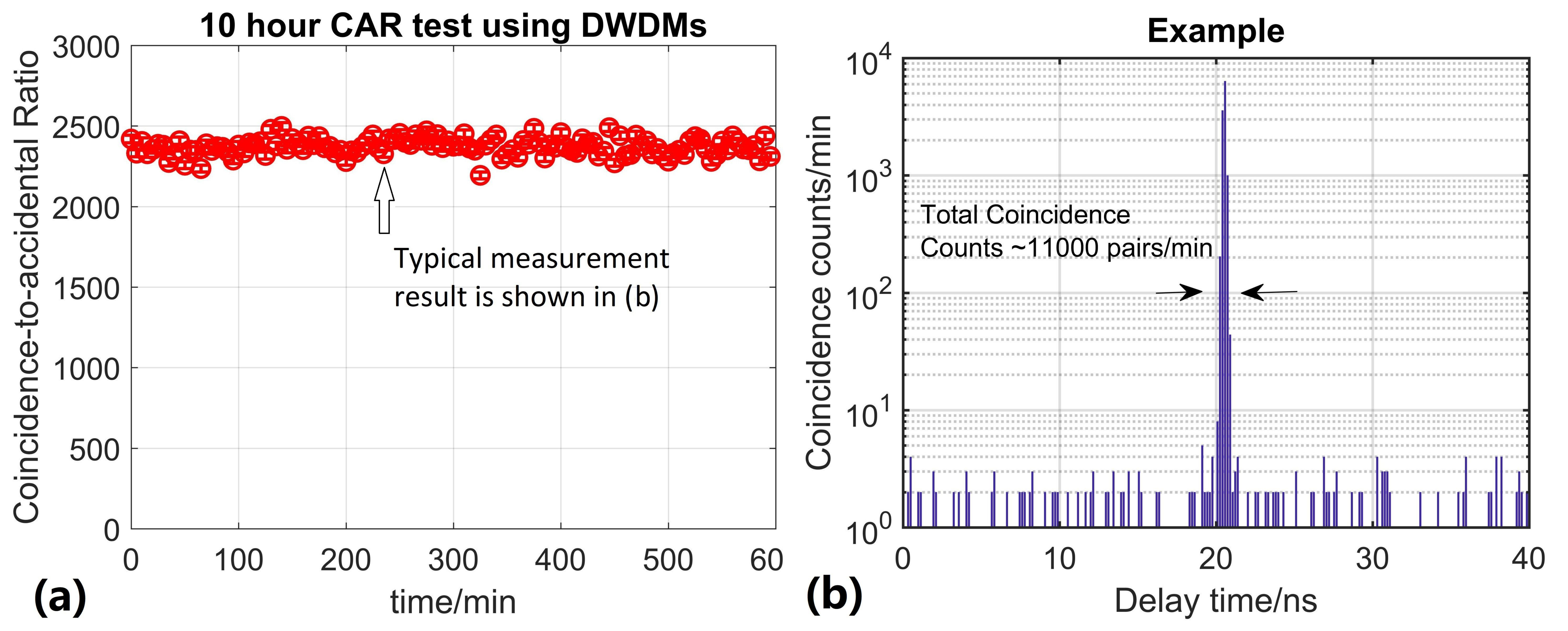}
\caption{(a) Coincidence-to-accidental ratio(CAR) in 10 hours using a DWDM filters set with effective bandwidth of 0.85nm, under a pump power of 7.5mW. (b) A typical coincidence example of 1min measurement histogram is randomly selected from (a). It demonstrates around 1.1$\times$10$^4$ pairs/min coincidence counts with around 3dB of losses after the output and 20$\%$ detection efficiency at single photon detectors.}
\label{fig:CAR}
\end{figure}

The stability of the entangled photon pairs source demands stable pump wavelength and stable phase-matching wavelength of PPSF. Since coincidence counts are more sensitive to  pump wavelength drift and phase-matching wavelength drift when narrower bandwidths of signal and idler filters are used, we used a pair of DWDM filters (passbands center at 1554.95nm and 1577.05nm with a 3dB bandwidth of 1.1nm) for demonstrating the operational stability of the source, as shown in Fig.\ref{fig:setup}(e). Using the narrower DWDM filters for signal / idler separation is more challenging than using the C/L band splitter when demonstrating coincidence-to-accidental ratio (CAR) as any wavelength drift (of pump, of PPSF, or of the filters) can significantly reduce the CAR. We continuously measured the CAR over a 10 hours period without any additional control of environmental factors at a pump power of 7.5mW. At an ambient temperature of 21.0$\pm$$2.0^o$C, the CAR maintained at approximately 2400, exhibiting excellent low-noise operational stability over the course of 10 hours, as shown in Fig.\ref{fig:CAR}(a). Note, the CAR would be even higher (estimated to be >3000) if we had used a pair of frequency-conjugate filters. A typical coincidence histogram is shown in Fig.\ref{fig:CAR}(b), indicating a detection rate of >200 pairs/nm/s (IDQ-220, deadtime 15$\mu$s, detection efficiency 20$\%$). The insertion loss in the DWDM filter sets is measured to be 3dB in total (1dB in WDM-1 transmission loss, 2 dB in the WDM-1 reflection and WDM-2 transmission loss, as shown in Fig.\ref{fig:setup}(e)) contributing to the reduction of coincidence. A total generation rate at the output of the source is thus estimated to be 7$\times$10$^5$pairs/s. The generation rate can be maximized up to 2.8$\times$10$^6$pairs/s by increasing the pump power at the expense of lowering CAR. 

The performance of the source can be further improved. The internal loss at present is largely due to large splice and component losses. For example, the PMF-PPSF and PPSF-SMF splices contribute $\sim$1.5dB loss each, while it can be further optimized to $<$0.5dB by using splicers with better heat control and distribution. And the pump suppression filters brings in another 4dB loss, which can be reduced by using better components.

In this work, we have demonstrated a compact, turn-key, polarization-entangled photon pair source based on the PPSF technology and incorporating a low-cost laser diode and other fiber-pigtailed telecom components. This source produces high entanglement quality (>98$\%$ fidelity) and broad bandwidth (>100nm), which can be a rich resource for quantum communications, quantum sensing, and quantum interferometry. Owing to the extremely simple, alignment-free structure, the source is reliable, robust and easy to use. It is an easily deployable polarization-entanglement source for a broad range of quantum applications, and further paves the way towards practical quantum technologies.

\section*{Acknowledgement}
C.C. would like to thank Olinka Bedroya and Xiaoqing Zhong for their experimental assistance. 

\section*{Funding}
Ontario Centres of Excellence VIP fund (project 26581); Natural Science and Engineering Research Council of Canada Engage fund (507623-17); OZoptics Ltd. 
\bibliography{sample}

\end{document}